\documentclass[graybox]{svmult}

\usepackage{mathptmx}       
\usepackage{helvet}         
\usepackage{courier}        
\usepackage{type1cm}        
\usepackage{makeidx}         
\usepackage{graphicx}        
\usepackage{multicol}        
\usepackage[bottom]{footmisc}

\makeindex      

\begin{document}

\title{Functional Cliques in Developmentally Correlated Neural Networks}
\titlerunning{Functional Cliques}
\author{S. Luccioli,  A. Barzilai, E. Ben-Jacob, P. Bonifazi, and A. Torcini}
\institute{Stefano Luccioli \at
CNR - Consiglio Nazionale delle Ricerche - Istituto dei Sistemi Complessi, 50019 Sesto Fiorentino, Italy;
INFN - Istituto Nazionale di Fisica Nucleare - Sezione di Firenze, 50019 Sesto Fiorentino, Italy: 
Joint Italian-Israeli Laboratory on Integrative Network Neuroscience, Tel Aviv University, Ramat Aviv, Israel.
\email{stefano.luccioli@fi.isc.cnr.it}
\\
\and Ari Barzilai \at Joint Italian-Israeli Laboratory on Integrative Network Neuroscience, Tel Aviv University, Ramat Aviv, Israel; Department of Neurobiology, George S. Wise Faculty of Life Sciences and Sagol School of Neuroscience, Tel Aviv University, Ramat Aviv, Israel.
 \\
\and Eshel Ben-Jacob \at 
Joint Italian-Israeli Laboratory on Integrative Network Neuroscience, Tel Aviv University, Ramat Aviv, Israel:
Beverly and Sackler Faculty of Exact Sciences School of Physics and Astronomy, Tel Aviv University, Ramat Aviv, Israel.
\and Paolo Bonifazi$^{(*)}$ \at Joint Italian-Israeli Laboratory on Integrative Network Neuroscience, Tel Aviv University, Ramat Aviv, Israel;Beverly and Sackler Faculty of Exact Sciences School of Physics and Astronomy, Tel Aviv University, Ramat Aviv, Israel;
Computational Neuroimaging Lab, BioCruces Health Research Institute, Hospital Universitario Cruces,
Plaza de Cruces, s/n E-48903, Barakaldo, Spain.
\email{paol.bonifazi@gmail.com}
\\
\and Alessandro Torcini$^{(*)}$\at 
Laboratoire de Physique Th\'eorique et Mod\'elisation
Universit\'e de Cergy-Pontoise - CNRS, UMR 8089,
95302 Cergy-Pontoise cedex, France;
Aix Marseille Univ, Inserm, INMED, Institute de Neurobiologie
de la M\'editerran\'ee and INS, Institut de Neurosciences des Syst{\' e}mes, Marseille, France;
Aix-Marseille Universit\'{e}, Universit\'{e} de Toulon, CNRS, CPT, UMR 7332, 13288 Marseille, France;
CNR - Consiglio Nazionale delle Ricerche - Istituto dei Sistemi Complessi, 50019 Sesto Fiorentino, Italy:
Joint Italian-Israeli Laboratory on Integrative Network Neuroscience, Tel Aviv University, Ramat Aviv, Israel.
\email{alessandro.torcini@u-cergy.fr}\\
$(*)$ These authors are joint senior authors on this work.
}
%

%
\maketitle



\abstract{We consider a sparse random network of excitatory leaky integrate-and-fire neurons
with short-term synaptic depression. Furthermore to mimic the dynamics of a brain circuit in 
its first stages of development we introduce for each neuron correlations among in-degree
and out-degree as well as among excitability and the corresponding total degree,
We analyze the influence of single neuron
stimulation and deletion on the collective dynamics of the network.
We show the existence of a small group of neurons capable of controlling 
and even silencing the bursting activity of the network. These neurons form
a {\it functional clique} since only their activation in a precise order and
within specific time windows is capable to ignite population bursts.
}

\section{Introduction}
\label{sec:1}

The relationship among brain functional and structural connectivity 
and neuronal  activity is one of the subject of major interest in neuroscience
~\cite{friston1994,Bullmore2009,Feldt2011,Bullmore2012,Lee2011}. Furthermore,
this research line is also strongly related to nonlinear dynamics themes regarding the influence of
topology on the emergence of specific microscopic and collective dynamics~\cite{bocca2006,olmi2010,luccioli2012}.
 
On one side, it is nowdays clear that a specific topology is not sufficient to
guarantee an unequivocal dynamical behaviour in the neural network~\cite{Gaiteri2011}.
On the other side, in the last two years experimental and numerical evidences have been indicating 
that neuronal ensembles or cell assemblies ({\it cliques}) are the emergent functional 
units of cortical activity~\cite{miller,seabrook,barnes,noi}, shaping  spontaneous 
and stimuli/task-evoked responses.  Interstingly,
several experimental studies have revealed that single  neurons can have a relevant role in shaping neuronal dynamics  in brain 
circuits~\cite{tsodyks1999,Brecht2004,Houweling2007,Cheng2009,Wolfe2010,London2010,Kwan2012,Bonifazi2009}.
Therefore it is of major importance to understand 
how do neuronal cliques emerge and how they do operate 
in relation to single neuron dynamics.
In this framework we could link the single neuron firing to 
the emergence of neuronal cliques, in a sort of hierarchical 
modular approach  where high order dynamics (i.e involving large number of neurons) 
can be strongly impacted by single neuron manipulations.

Following this line of experimental and theoretical evidence, 
and in particular the analysis performed in~\cite{Bonifazi2009},
we have derived a numerical model displaying collective oscillations, similar to the ones
observed in the hyppocampus and in the neocortex at the early stage of development~\cite{Allene2008}.
In particular, we have analyzed the influence of single neuron stimulation/deletion on the
dynamics of such a network in presence of developmentally inspired constraints. Our
analysis reveals that a small group of critical neurons, organized in a {\it clique}, is capable to control the bursting behaviour  of the whole network. These neurons are {\it hubs} in a {\it functional} sense, because their
relevance is not related to high intrinsic connectivity, but to their precise sequential and coordinated 
activation before a bursting event. As a consequence, if a perturbation is applied to any 
critical neuron of the {\it functional clique},  through their stimulation or the deletion, 
their sequential activation can be interrupted, thus leading to dramatic consequence
at a network level, with the arrest of the collective oscillations.

The studied model and the methods employed to analyze the data
are introduced in Sect.~\ref{sec:2}, with particular emphasis on functional
connectivity.  Sect.~\ref{sec:3} reports the results of the numerical experiments
performed at the single neuron level in our network and of the functional analysis.
Finally a brief discussion is reported in Sect.~\ref{sec:4}.

\section{Model and Methods}
\label{sec:2}

We consider a network of $N$ excitatory Leaky Integrate-and-fire (LIF) neurons,
interacting via synaptic currents regulated by short-term-plasticity
according to the model introduced in \cite{Tsodyks2000Synchrony}.
For this model, the evolution of the membrane potential $V_i$ of the neuron $i$ is
given by
\begin{equation}
\label{eq1n}
\tau_{\mathrm{m}} \dot V_i= -V_i + I^{\mathrm{syn}}_{i}+I^{\mathrm{b}}_{i} \, 
\end{equation}
where $\tau_\mathrm{m}$ is the membrane time constant, 
$I^{\mathrm{syn}}_{i}$ is the synaptic current received by neuron $i$ from 
all its presynaptic inputs and $I^{\mathrm{b}}_{i}$ represents its level of 
intrinsic excitability, encompassing single neuron properties as well as 
excitatory inputs arriving from distal areas.The currents are measured in voltage units (mV),
since the membrane input resistance is incorporated into the currents.

Whenever the membrane potential $V_i(t)$ reaches the threshold 
value $V_{\mathrm{th}}$, it is reset to $V_{\mathrm{r}} $, and a $\delta$-spike is sent 
towards the postsynaptic neurons. 
Accordingly, the spike-train $S_j(t)$ produced by neuron $j$, is defined as,
\begin{equation}
S_j(t)=\sum_m \delta(t-t_{j}(m)),
\end{equation}
where $t_{j}(m)$ represent the $m$-th spike time emission of neuron $j$.
The transmission of the spike train $S_j$ to the efferent neurons is mediated by 
the synaptic evolution. In particular, by following \cite{Tsodyks1997} 
the state of the synapse between the $j$th presynaptic neuron and the $i$th postsynaptic neuron is described by three adimensional variables, $X_{ij}$, $Y_{ij}$, and $Z_{ij}$, 
representing the fractions of synaptic resources in the recovered, active, and inactive state, 
respectively. 

Since these three variables should satisfy the constraint 
$X_{ij}+Y_{ij}+Z_{ij}=1$, it is sufficient to provide the time
evolution for two of them, namely
\begin{equation}
\label{dynsyn}
\dot Y_{ij} = -\frac{Y_{ij}}{T^I_{ij}} +u_{ij}X_{ij}S_{j}
\end{equation}
\begin{equation}
\label{contz}
\dot Z_{ij} = \frac{Y_{ij}}{T^\mathrm{I}_{ij}}  -   \frac{Z_{ij}}{T^\mathrm{R}_{ij}} .
\end{equation} 
Only the active transmitters react to the incoming spikes $S_j$
and the adimensional parameters $u_{ij}$ tunes their effectiveness. 
The decay times of the postsynaptic current are given by $\{T^\mathrm{I}_{ij}\}$,
while $\{T^\mathrm{R}_{ij}\}$ represent the recovery times from the synaptic depression. 

The synaptic current is expressed as the sum 
of all the active transmitters (post-synaptic currents) delivered to neuron $i$ 
\begin{equation}
\label{curr}
I^{\mathrm{syn}}_{i} = \frac{ G_{i}}{K^{I}_{i}}\sum_{j\ne i}  \epsilon_{ij}Y_{ij},
\end{equation}
where $G_{i}$ is the coupling strength, while $\epsilon_{ij}$ is the connectivity
matrix whose entries are set equal to $1$ ($0$) if the presynaptic neuron
$j$ is connected to (disconnected from) the postsynaptic neuron $i$. At variance 
with~\cite{Tsodyks2000Synchrony}, we assume that the coupling strengths are the same for all the
synapses afferent to a certain neuron $i$. 
 
In particular, we study the case of excitatory coupling between neurons,
i.e. $G_{i} > 0$.  Moreover, we consider a sparse network made of $N=200$ neurons where the 
$i$-th neuron has $K^{I}_{i}$ ($K^{O}_{i}$) afferent (efferent) synaptic connections 
distributed as in a directed Erd\"os-R\'enyi graph  with average in-degree $\bar K^I =10$,
as a matter of fact also the average out-degree was $\bar K^0 =10$.
The sum appearing in (\ref{curr}) is normalized by the input degree $K^I_{i}$ 
to ensure homeostatic synaptic inputs~\cite{Turrigiano1998,Turrigiano2008}.

The intrinsic excitabilities of the single neurons $\{I^b_{i}\}$ are randomly chosen from a 
flat distribution of width 0.45 mV centered around the value $V_{\mathrm{th}} = 15$ mV, by
further imposing that 5\% of neurons are above threshold in order to observe a bursting
behaviour in the network. For the other parameters, in analogy with Ref.~\cite{Tsodyks2000Synchrony},
we use the following set of values: $\tau_\mathrm{m} = 30$ ms,  
$V_{\mathrm{r}} = 13.5$ mV, $V_{\mathrm{th}} = 15$ mV. The synaptic 
parameters $\{T^\mathrm{I}_{ij}\}$, $\{T^\mathrm{R}_{ij}\}$, $\{u_{ij}\}$ and $\{G_{i}\}$ are
Gaussian distributed with averages $\overline{T^\mathrm{I}} = 3$ ms, 
$\overline{T^\mathrm{R}} = 800$ ms, $\overline{u} = 0.5$ and $\overline{G} = 45$ mV, 
respectively, and with standard deviation equal to the half of the average.

\subsection{Correlations}

In this paper, we intend to mimic neural networks
at the first stage of development. In such networks
mature and young neurons are present at the same time, and
this is reflected in the variability of the structural connectivities and of the
intrinsic excitabilities. In particular, experimental observations indicate that
younger cells have a more pronounced excitability~\cite{Ge2005,Doetsch2005},
while mature cells exhibit a higher number of synaptic inputs~\cite{Bonifazi2009, Marissal2012}.
Thus suggesting that the number of afferent and efferent synaptic connections~\cite{Bonifazi2009,Marissal2012,Picardo2011}
as well as their level of hyperpolarization~\cite{Karayannis2012} are positively correlated with
the maturation stage of the cells. Therefore, we consider a network
including the following two types of correlations:
\begin{itemize}

\item{setup T1:} a positive correlation between the in-degree $K^{I}_{i}$
and out-degree  $K^{O}_{i}$ of each neuron;

\item{setup T2:} a negative correlation between the intrinsic neuronal excitability and the
total connectivity $K^T_i =K^{I}_{i} + K^{O}_{i}$ (in-degree plus out-degree) ;

\end{itemize}

Correlation of type T1 is obtained by generating randomly two pools of $N$ input and output degrees 
from an Erd\"os-R\'enyi distribution with average degree equal to 10. The degrees are ordered
within each pool and then assigned to $N$ neurons in order to obtain a positive correlation 
between $K^{O}_{i}$ and $K^{I}_{i}$. 

Correlation of type T2 imposes a negative correlation between
excitability $I^b_{i}$ and the total degree of the single neuron $K^T_i =K^{I}_{i} + K^{O}_{i}$.
To generate this kind of correlation the intrinsic excitabilities are randomly generated,
as explained above, and then assigned to the various neurons 
accordingly to their total connectivities $K_i^T$, thus to ensure an inverse correlation 
between $I^b_{i}$ and $K_i^T$. 
 
\subsection{Functional Connectivity}

In order to highlight statistically significant time-lagged activations of neurons, 
for every possible neuronal pair, 
we measure the cross-correlation between their spike time series. On the basis of 
this cross-correlation we eventually assign a directed functional 
connection among the two considered neurons, similarly to what reported 
in \cite{Bonifazi2009, Bonifazi2013} for calcium imaging studies.

For every neuron, the action potentials timestamps were first converted into a binary 
time series with one millisecond time resolution, where ones (zeros) marked the occurrence 
(absence) of the action potentials. Given the binary time series of two neurons $a$ and $b$, the 
cross correlation was then calculated as follows:
\begin{equation}
C_{ab} (\tau) = \frac{\sum_{t=\tau}^{T-\tau} a_{t+\tau} b_t}{min(\sum_{i=1}^T a_i, \sum_{k=1}^T b_k)} 
\end{equation}
where $\{a_t\}$,$\{b_t\}$ represented the considered time series and $T$ was their total duration.
Whenever $C_{ab} (\tau)$ presented a maximum at some finite time value $\tau_{max}$ a functional
connection was assigned between the two neurons: for $\tau_{max} < 0$ ($\tau_{max} > 0$) 
directed from $a$ to $b$ (from $b$ to $a$). A directed functional connection cannot be defined
for an uniform cross-correlation corresponding to uncorrelated neurons or for synchronous firing
of the two neurons associated to a Gaussian $C_{ab} (\tau)$ centered at zero.
To exclude the possibility that the cross correlation could be described by a Gaussian with zero mean 
or by a uniform distribution we employed both the Student\'\-s t-test and the Kolmogorov-Smirnov test with 
a level of confidence of 5\%. The functional out-degree $D^O_i$ (in-degree $D^I_i$) of a neuron $i$ 
corresponded to the number of neurons which were reliably activated after (before) its firing.

For more deiails on the model and methods employed to perform the reported analysis see ~\cite{noi}.

\section{Results}
\label{sec:3}

\begin{figure}[h]
\includegraphics[scale=.24]{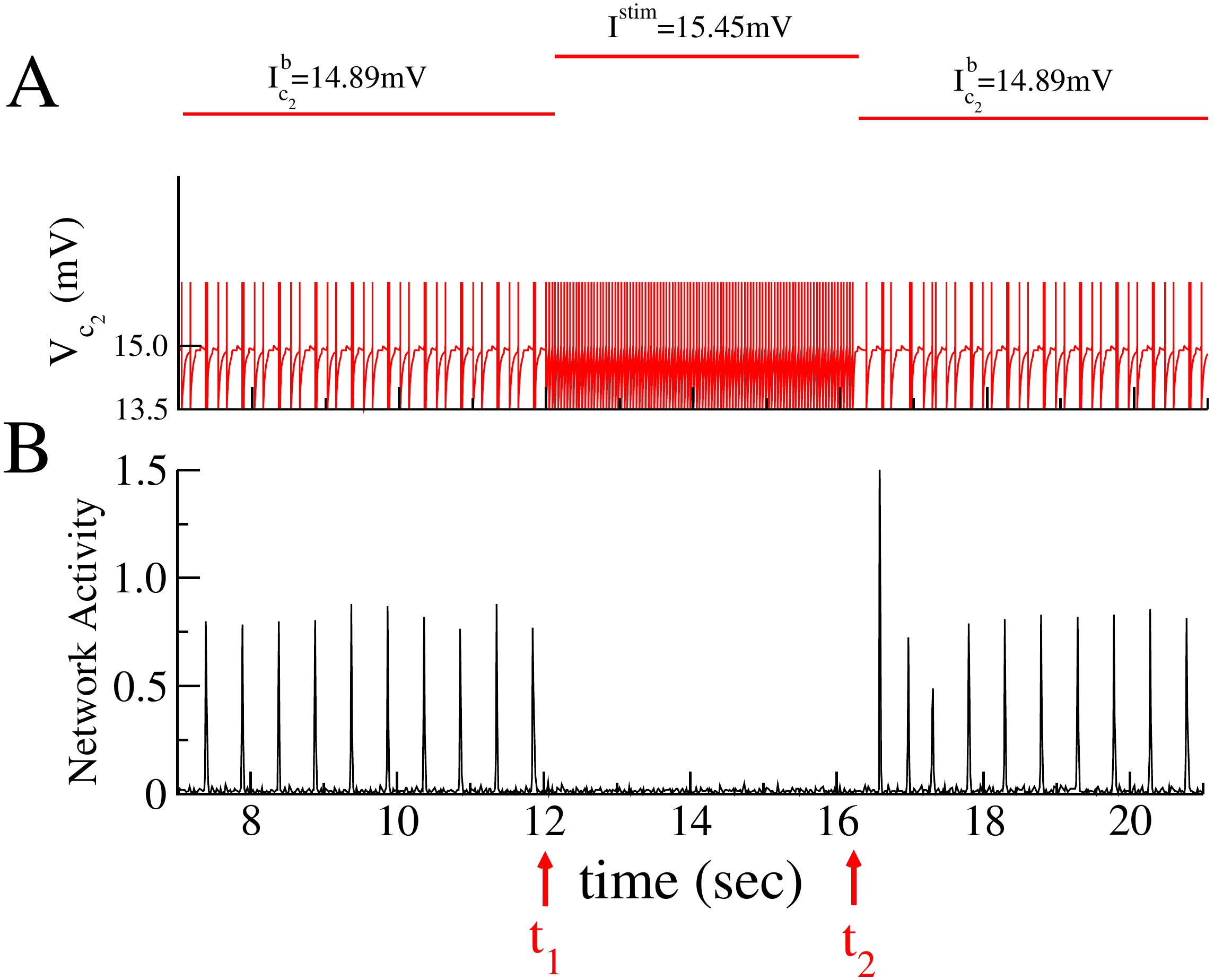}
\includegraphics[scale=.24]{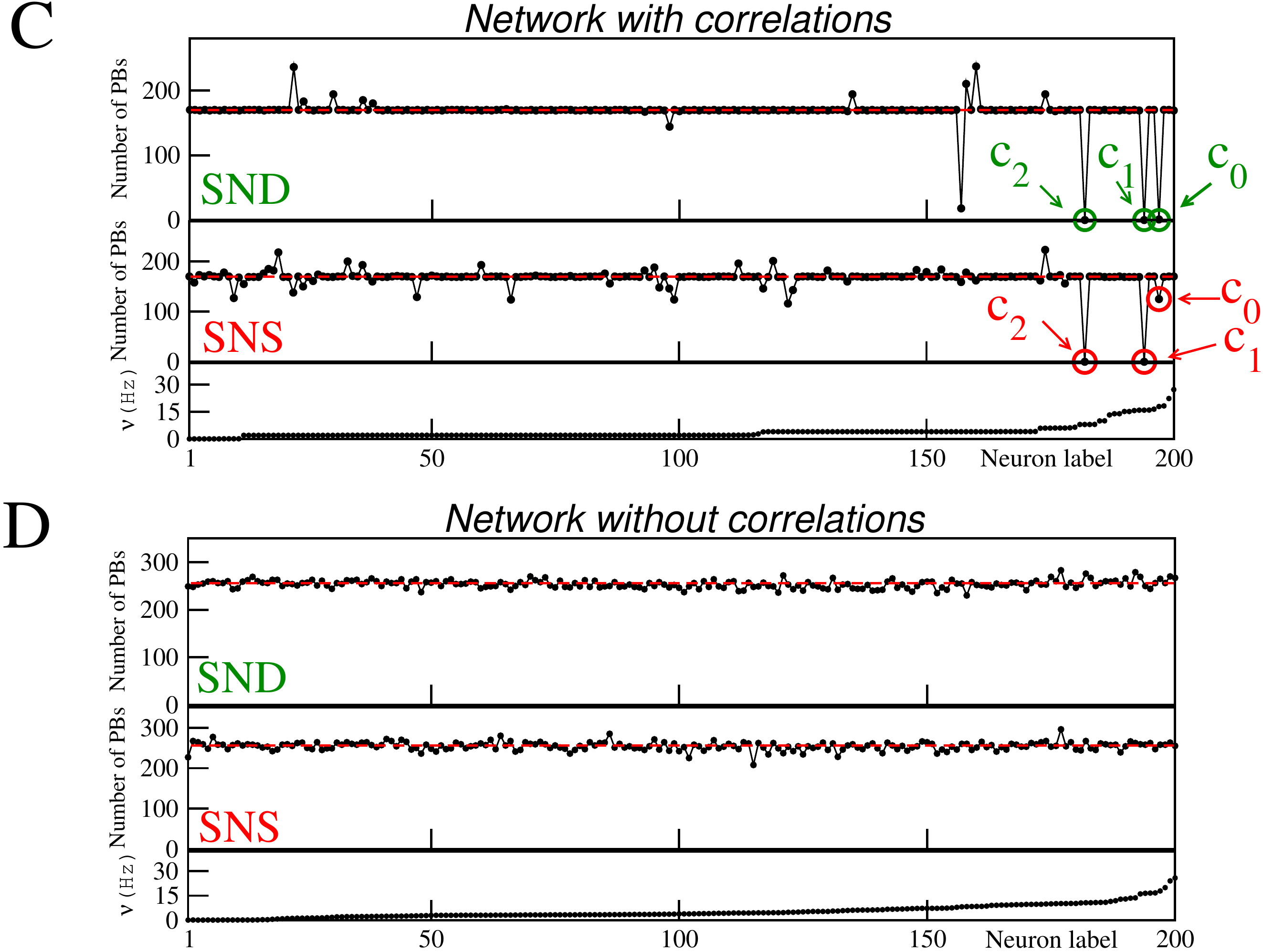}
%
%
\caption{(A), (B) Sketch of a SNS experiment for a network with type 
T1 plus T2 correlations. The neuron $c_2$ is stimulated with a DC step for a time 
interval $\Delta t=t_{2}-t_{1}$ (see the red line on the top panel). 
In (A) the membrane potential $V_{c_2}$ of the stimulated neuron is shown 
(firing times are signaled by vertical bars of height 1.5 mV), while in (B) the network activity 
is reported. (C) and (D) refer to correlated and uncorrelated networks,
respectively, and display the number of PBs emitted
in a time window $\Delta t = 84$ s during SND and SNS experiments with $I^{{\rm stim}}=15.45$ mV. 
The horizontal dashed lines refer to the
average number of PBs emitted in a time interval $\Delta t$ = 84 s during a control experiment when no
stimulation is applied (the amplitude of the fluctuations is smaller than the symbols).
Neurons are ordered accordingly to their average firing rates $\nu$ measured in 
control condition (plotted in the bottom panels). 
The critical neurons $c_0$, $c_1$, $c_2$, signaled by green (red) circles, 
are able to strongly affect the bursting during SND (SNS). During SNS experiments 
each neuron $i$ was stimulated with a DC step switching its excitability 
from $I_{i}^{b}$ to $I^{{\rm stim}}$ for an interval $\Delta t= 84$ s. 
The data refer to $I^{{\rm stim}}=15.45$ mV and $N=200$ neurons.}
\label{fig:1}       
\end{figure}

In this paper we intend to mimic an immature neuronal network 
at a certain stage of its initial development, in analogy
with the one examined in the experimental work on developmental 
hippocampal circuits~\cite{Bonifazi2009}. As discussed in~\cite{Bonifazi2009,Allene2008},
at early postnatal stages such networks are characterized by 
the excitatory action of GABAergic transmission and the presence of synchronized 
network events, as largely documented in central and peripheral nervous circuits.

Therefore, we consider a network model composed of only excitatory neurons
and displaying bursting activity. A minimal model  to mimic the
experimentally described stereotypical/characteristic condition of developing neuronal networks~\cite{Ben2002} is the one introduced by Tsodyks-Uziel-Markram (TUM) \cite{Tsodyks2000Synchrony}
with purely excitatory synapses. Since this model (described in details in Sect.~\ref{sec:2})
is known to display an alternance of short periods of quasi-synchronous firing
({\it population bursts}, PBs) and long time intervals of asynchronous firing.
Furthermore, we consider a network with embedded correlations of type T1 and T2,
this in order to account for the 
presence at the same time of younger and older neurons. 
This presence can be modeled by considering 
correlations among the in-degree and out-degree of each cell (type T1)
as well as among their intrinsic excitability and connectivity (type T2),
as already explained in the previous Section.
The network activity was characterized by bursts of duration $\simeq 24$ ms
and with interburst intervals $\simeq 500$ ms.

\subsection{Single neuron stimulation and deletion experiments}

In the developing hippocampus it has been shown that the stimulation of specific single neurons
can drastically reduce the frequency of the PBs~\cite{Feldt2011,Bonifazi2009},
or even silence the collective activity. More specifically,
the stimulation consisted of current injection capable of inducing sustained 
high firing regime of the stimulated neuron over a period of a few minutes. 

Inspired by this experimental protocol, we test the impact of {\it single neuron stimulation} (SNS)
on the occurrence of PBs in our network model.
SNS was achieved by adding abruptly a DC current term to the considered neuron.
We report in Fig.~\ref{fig:1} A-B the stimulation protocol for a specific neuron 
capable of suppressing the occurrence of PBs for all the duration of the SNS 
(in this case limited to $4.2$ s). The SNS process is totally reversible, i.e. 
when the stimulation is interrupted the firing rate
of the cell and the PBs frequency returns to the pre-stimulation control level.

\begin{figure}[h]
\includegraphics[scale=.4]{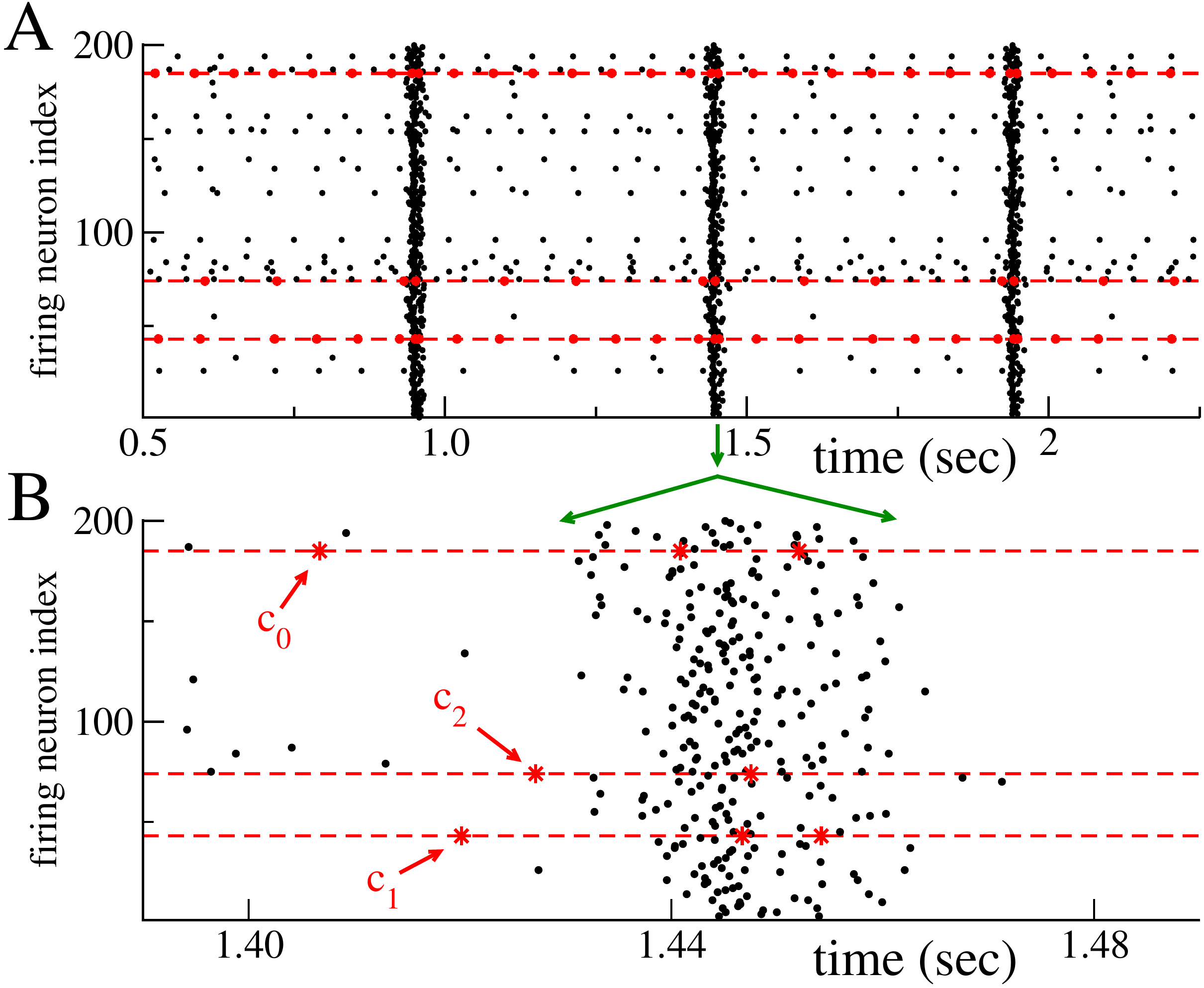}
%
%
\caption{(A) Raster plot of the network activity: every dot signals a firing event. The
(red) dashed lines and (red) stars refer to the critical neurons. (B) Close up of a 
population burst: PBs are anticipated by the ordered firing sequence of the critical neurons 
$c_0 \rightarrow c_1 \rightarrow c_2$. 
In the raster plots, at variance with all the other figures, 
the neuronal labels are not ordered accordingly the ascending firing rates.}
\label{fig:3}       
\end{figure}

In order to measure the impact of SNS on the network dynamics, we consider
the relative variation of the PB frequency 
with respect to the control conditions (i.e. in absence of any stimulation).
In Figs.~\ref{fig:1} C-D the impact of a SNS on the
PBs frequency is reported for a classical 
Erd\"os-R\'enyi network (no correlations) and a network with embedded correlations
T1 plus T2. It is clear that SNS has much more impact on the correlated network, than in the
uncorrelated one. In particular for two neurons, indicated as neuron
$c_1$ and $c_2$, SNS performed 
with stimulation current $I^\mathrm{stim}=15.45$ mV, 
is able to suppress the occurrence of PBs during the stimulation
period. Furthermore in other 7 cases SNS reduces the collective activity $\simeq 30\%$,
in the specific case of neuron $c_0$ (whose relevance is discussed in the following)
the reduction was of $\simeq 27\%$.

On the contrary,
in the network without correlations, the SNS had only marginal effect on the
population activity, although the distributions of the firing rates
in the correlated and uncorrelated network are extremely similar (under control conditions)
as shown in the bottom panels of Fig.~\ref{fig:1} C and D.

In \cite{Tsodyks2000Synchrony} it has been shown that the elimination of a pool of neurons from an
uncorrelated TUM network induced a strong reduction of the population bursts. 
In our analysis we repeat such numerical experiment with single cell resolution, i.e.
we consider the influence of {\it single neuron deletion}, SND, on the 
network response. The results reported in Fig.~\ref{fig:1} C-D,
clearly show almost no effect for the uncorrelated network.
However, SND blocks the deliver of PBs in the network with correlations $T1$ plus $T2$, 
in three cases. It should be noticed that two over three neurons were critical also fro the SNS,
while the third critical neuron has a higher frequency and it is denoted as $c_0$.

The most critical neurons are all unable to fire if isolated from the network,
i.e. they are all below threshold, apart neuron $c_0$, which has $I^b_{c_{0}} = 15.19$ mV.
Therefore we observe that SND or SNS are capable to silence
the network and that this occurs only for a quite limited number of neurons,
which have reasonably high $I^b$ but low $K^T$ 
($c_0$, with $K^T$=6, is ranked as the second less connected neuron of the network).

\begin{figure}[h]
\includegraphics[scale=.28]{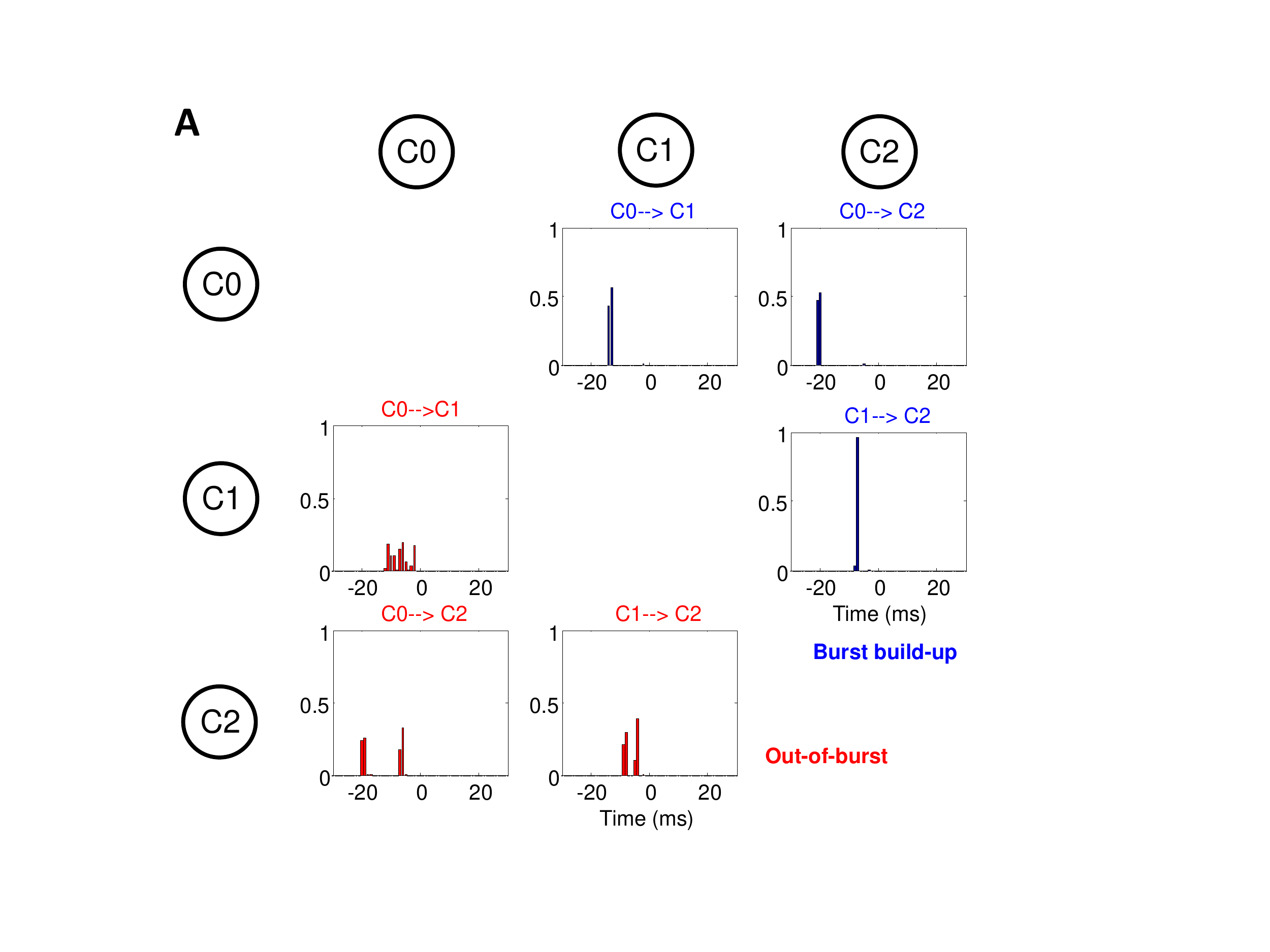}
\hspace{-1.0cm}
\includegraphics[scale=.28]{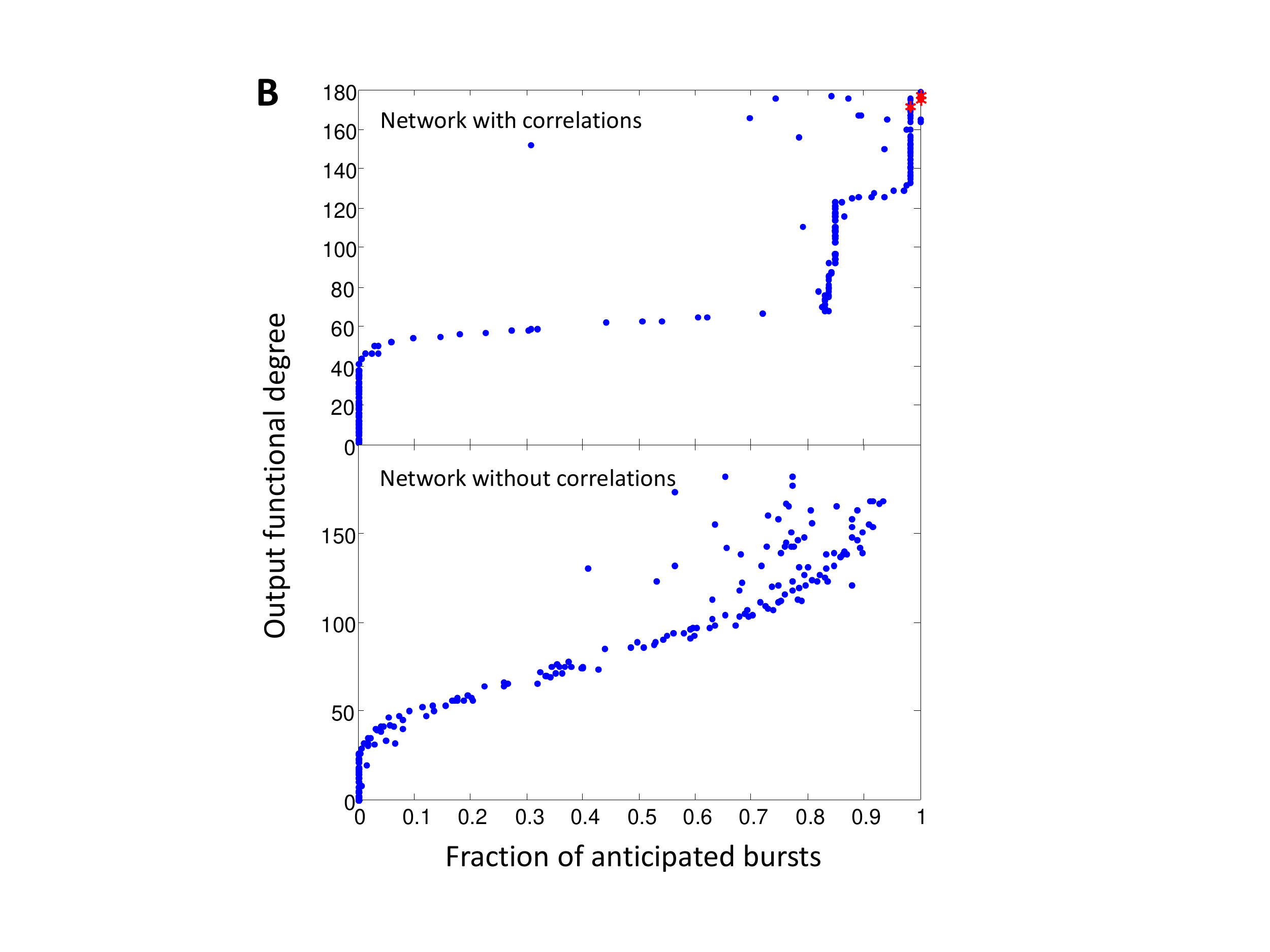}
%
%
\caption{(A) Cross correlation functions $C(\tau)$ between the spike trains of two critical neurons for the network with correlations. The plots show all the possible
pair combinations of the critical neurons: blue (red) histograms 
refer to the analysis performed during the population burst build up 
(during periods out of the bursting activity). The order of activation of each pair is reported on the
top of the corresponding plot, whenever the cross-correlation has a significant maximum
at some finite time $\tau_{max}$. (B) Output functional degree as a function of 
the fraction of the anticipated population bursts for all the neurons in the network:
each dot denotes a different neuron. The data in the top (bottom) panel 
refer to the network with correlations (without correlations).
The (red) stars on the top right in the top panel signal the critical neurons $c_{0},c_{1},c_{2}$.}
\label{fig:2}       
\end{figure}

\subsection{The clique of functional hubs}

\begin{figure}[h]
\includegraphics[scale=.4]{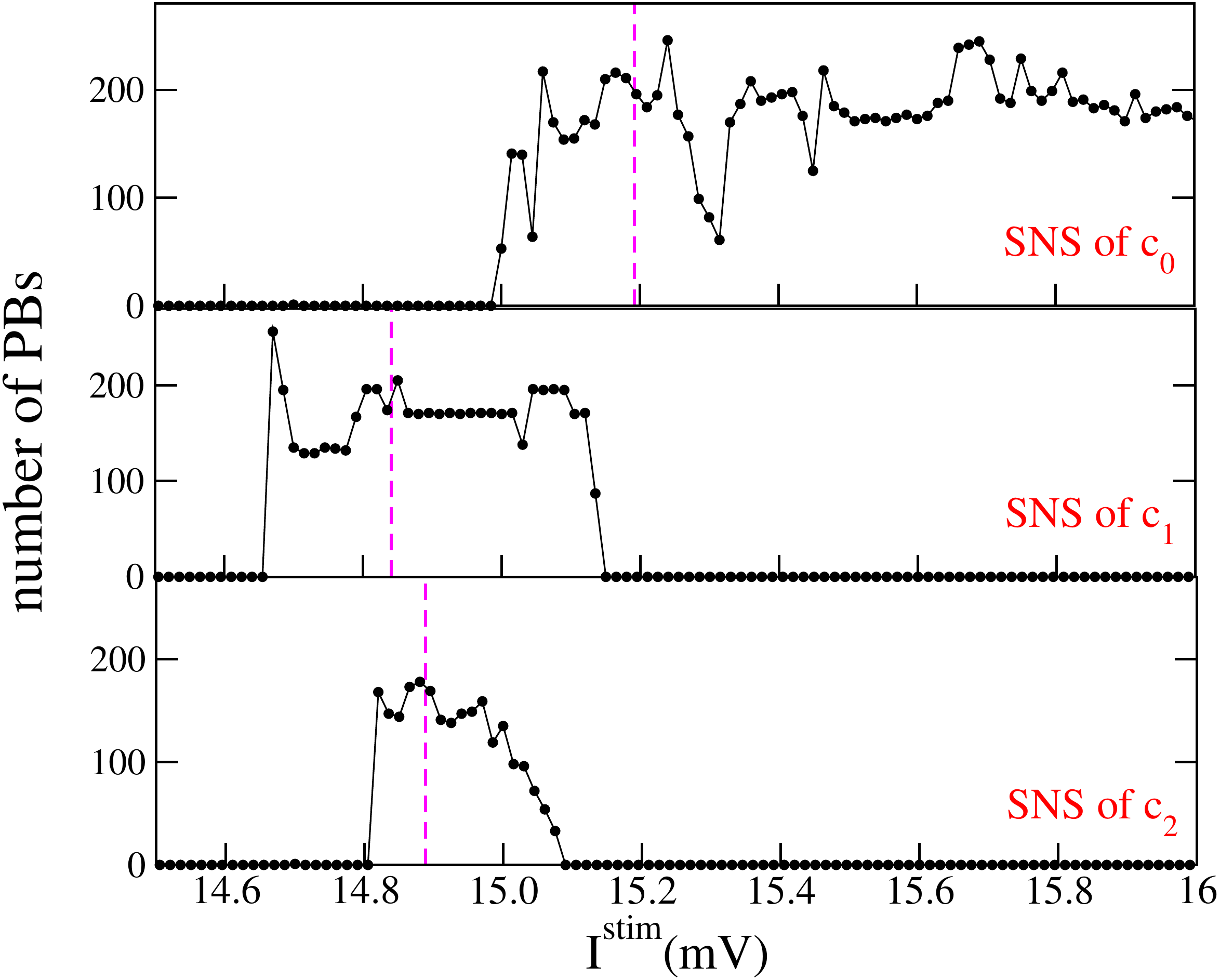}
%
%
\caption{Number of PBs emitted in a time window $\Delta t = 84$ s during SNS experiments of 
the critical neurons $c_0$, $c_1$, $c_2$ versus the stimulation current $I^{{\rm stim}}$. 
The vertical dashed lines signal the value of the intrinsic excitability in control condition.}
\label{fig:4}       
\end{figure}

The detailed investigation of the burst events revealed that each PB is
always preceeded by the firing of the 3 critical neurons in the following order: 
$c_0 \to c_1 \to c_2  \to$ PB, as it is shown in Fig.~\ref{fig:3}.
The neuron $c_0$, which is the only one supra-threshold, fires first 
followed by the others (sub-threshold) and this trigger the onset of the PB.
Furthermore, the functional connectivity analysis of the neurons
in the network confirms that $c_0$, $c_1$, and $c_2$ have quite high
functional output degree $D^{O} \simeq 170-180$ and that they essentially anticipate 
all the  PBs in the network, as shown in Fig.~\ref{fig:2} B. 
In more details, one observes that besides firing in a precise order
the three critical neurons always fire within a narrow temporal window
one after the other before a PB as shown in Fig.~\ref{fig:2} A,
(in particular see the blue cross correlation functions).
Namely, $c_1$ fires after $c_0$ with an average time delay $\Delta T_{c_1,c_0} \simeq 13.4$ ms, and then $c_2$ after $c_1$ within the average interval $\Delta T_{c_2,c_1} \simeq 7.0$ ms. 
Furthermore, the analysis of the activity of these three neurons
during the inter-bursts reveal that they do not show anymore an univoque
firing order or an unique activation time window (see in Fig.~\ref{fig:2} A the 
red cross correlations functions).

An extensive investigation of the critical neurons subjected 
to stimulations with currents in the range $I^\mathrm{stim} \in [14.5:16.0]$ mV
reveals that PBs can be observed only if the neurons $c_1$ and $c_2$ had
excitabilities within a narrow range (of amplitude $\simeq 0.2$ mV) centered around 
the threshold value. While, the stimulation of neuron 
$c_0$ reveals that the network is always active for $I^\mathrm{stim} > V_{{\rm th}}$, apart
the occurence of an absolute minimum in the PB activity
(an {\it anti-resonance}) at $I^\mathrm{stim} = 15.32$ mV and a relative minimum at 
$I^\mathrm{stim} = 15.45$ mV (see Fig.~\ref{fig:4}).

As shown in Fig.\ref{fig:4}, whenever $c_1$ and $c_{2}$ are stimulated with currents 
$I^{stim} > I^b_{c_{0}} = 15.19$ mV the bursting activity stops. 
This behaviour indicates that neuron $c_{0}$ is the {\it leader} of the clique 
and the other ones are simply {\it followers} in the 
construction of the PB, they cannot fire more rapidly then neuron 
$c_{0}$ or the PBs ceases. An analysis of the structural connectivity 
reveales that neuron $c_0$ projected an efferent synapse on $c_1$, 
which projected on $c_2$. 

The critical neurons, as already mentioned, are not hubs in a structural sense,
since they have a very low connectivity $K^T$, however they are indeed functional hubs, as shown from
the previous analysis. Therefore, we can 
affirm that neurons $c_0$, $c_1$ and $c_2$ form a functional clique, 
whose sequential activation, within the previously reported time windows,
triggers the population burst onset. 

The functional relevance of the neurons for the network dynamics is even more evident in other 
examples of functional cliques, reported in \cite{noi},  where some of the  supra-threshold neurons 
were even  not structurally connected among each other.

\section{Discussion}
\label{sec:4}

We have shown how, in a simple neural circuit displaying collective oscillations,
the inclusion of  developmentally inspired correlations among the single neuron excitabilities
and connectivities can lead to the emergence of a {\it functional clique}. This is a small
group of functionally connected neurons whose activation can be crucial to promote/arrest the collective 
firing activity in neural networks, irrespective of the underlying network topology. 
The clique is composed of a leader neuron, which can start the activation sequence at
any moment, but to ignite the population burst the other two neurons, the followers,
should fire in a precise order and with quite defined delays.

These results besides being of extreme interest for the neuroscience and dynamical system communities,
pave the way for a new approach to the control of the dynamics of neural circuits.
Coherence or decoherence in the network activity can be induced by proper stimulation
protocols of few peculiar neurons, a subject of extreme interest in particular for
the treatment of Parkinson disease with adaptive deep brain stimulation~\cite{littel2013}.

Future developments towards a more realistic neural circuit would require the extension of the model 
to include inhibitory neurons, as well as facilitation mechanisms at the level of
synaptic transmission, and of network topology arising from anatomical data.

\begin{acknowledgement}
We thank Y. Ben-Ari, D. Angulo-Garcia, R. Cossart, A. Malvache, 
L. M\'odol-Vidal, for extremely useful interactions.
This article is part of the research activity of the Advanced Study Group 2016
{\it From Microscopic to Collective Dynamics in Neural Circuits} performed
at Max Planck Institute for the Physics of Complex Systems in
Dresden (Germany).
\end{acknowledgement}
%

%
%
%

\biblstarthook{
\textcolor{red}{References may be \textit{cited} in the text either by number (preferred) or by author/year.
\footnote{\textcolor{red}{Make sure that all references from the list are cited in the text. 
Those not cited should be moved to a separate \textit{Further Reading} section or chapter.}} 
The reference list should ideally be \textit{sorted} in alphabetical order -- even if reference numbers are used for the their citation in the text. If there are several works by the same author, the following order should be used: 
\begin{enumerate}
\item all works by the author alone, ordered chronologically by year of publication
\item all works by the author with a coauthor, ordered alphabetically by coauthor
\item all works by the author with several coauthors, ordered chronologically by year of publication.
\end{enumerate}
}
}

\end{document}